# Effect of Li doping on structure and superconducting transition temperature of $Mg_{1-x}Li_xB_2$

Y. G. Zhao[1], X. P. Zhang[1], P. T. Qiao[1], H. T. Zhang[1], S. L. Jia[2], B. S. Cao[1]
M. H. Zhu[1], Z. H. Han[1], X. L. Wang[3], B. L. Gu[1]

[1] Department of Physics, Tsinghua University, Beijing 100084, P. R. China
[2] National Lab for Superconductivity, Institute of Physics, Chinese Academy of Sciences, Beijing 100080, P. R. China
[3] Functional Materials Institute, Central Iron & Steel Research Institute, Beijing 100081, P. R. China

We report the preparation of $Mg_{1-x}Li_xB_2$ compounds. Nearly single phased samples were obtained for $x \le 0.3$. The in-plane lattice parameter *a* decreases with Li doping, while the lattice parameter *c* does not show obvious change. The superconducting transition temperature of $Mg_{1-x}Li_xB_2$ decreases with Li doping and loss of superconductivity occurs for x=0.5 sample. The results of our work are consistent with the prediction of the hole superconductivity mechanism.

Very recently the surprising discovery of superconductivity in the diboride, $MgB_2$, has resulted in a flurry of both experimental and theoretical work [1]. This binary intermetallic superconductor shows superconductivity at 39 K, which is almost two times of the superconducting transition temperature ($T_c$) of the old record for the intermetallic superconductors. A significant boron isotope effect was observed in $MgB_2$ [2], suggesting a phonon mediated BCS superconducting mechanism in this compound [3, 4]. However, if BCS theory applies, $T_c$ of $MgB_2$ almost reaches the limit of $T_c$ of the theory. On the other hand, tt is proposed that the theory of hole superconductivity may also account for the superconductivity in $MgB_2$ [5, 6]. So the mechanism for the occurrence of superconductivity in $MgB_2$ is still an open question. It has been proposed that $Mg_{1-x}Li_xB_2$ is a key compound to test these two theories because of their different predictions on the behavior of $T_c$ upon hole doping [6].

In this paper, we report the preparation of $Mg_{1-x}Li_xB_2$ compounds and their structure and $T_c$ dependence on Li doping. The results show that nearly single phase $Mg_{1-x}Li_xB_2$ can be obtained for x at least up to 0.3 and multiphases coexist for $x \ge 0.5$ samples. Upon Li doping, the in-plane lattice parameter *a* decreases, while the lattice parameter *c* remains unchanged. The superconducting transition temperature decreases with Li doping. The results of our work favor the hole superconductivity model.

$Mg_{1-x}Li_xB_2$ samples with x=0, 0.1, 0.3, 0.5 and 1 were prepared by solid state reaction method. The starting materials were Mg flakes (99.9% pure), Li (99.99% pure in lump form) and amorphous B powder (99.99% pure) and they are combined in a sealed Ta tube in a stoichiometric ratio in a Ar environment. The Ta tube is then sealed in a quartz ampoule, placed in a box furnace and heated at 950 $^{o}$C for two hours, then quenched to room temperature. The phase analysis of the samples was performed using Rigaku D/max-RB x-ray diffractometer with Cu $k_\alpha$ radiation. AC susceptibility of the samples was measured from room temperature to liquid helium temperature.

The x-ray diffraction patterns for x=0, 0.1 and 0.3 are very similar. Fig. 1 is the x-ray diffraction pattern for x=0.1 sample which is consistent with the reported x-ray diffraction pattern of $MgB_2$ [2] indicating that the sample is nearly single phased. The peak around $2\theta$=38.51 also exists in the undoped $MgB_2$ [2]. The x-ray diffraction pattern for x=0.5 samples, however, shows complicated features, quite different from the x-ray diffraction patterns for $x\leq0.3$ samples and may contain multiphases in the samples. Although the x-ray diffraction patterns for x=0, 0.1 and 0.3 samples are similar, the positions of the diffraction peaks for (100), (101) and (110) show dramatic increase with doping, especially the (100) and (110) peaks, while the position of (002) peak remains unchanged. Fig. 2 shows the variation of peak positions for (110) peaks with Li doping. This result indicates that the in-plane lattice parameter *a* decreases with Li doping and the lattice parameter *c*, which is the distance between the adjacent B layers, remains unchanged within the experimental error. Therefore Li doping mainly affects the in-plane coupling. Fig. 3 is the variation of lattice parameters *a* and *c* with Li doping. It can be seen that the lattice parameter a decrease dramatically with Li doping while lattice parameter *c* does not show systematic change. The contraction of the unit cell of $Mg_{1-x}Li_xB_2$ can be understood by considering the fact that the ion radius of $Li^+$ (0.6 Å) is smaller than that of $Mg^{2+}$ (0.65 Å).

Fig. 4 shows the temperature dependence of AC susceptibility for $Mg_{1-x}Li_xB_2$ with different Li doping. For x=0.1 sample, $T_c$ does not show obvious change as compared to that of the undoped sample. This may be due to a combined effect of the mass reduction on Mg site and other effects induced by Li doping. The former increases $T_c$, while the latter may decrease $T_c$. This phenomenon deserves further study. The significance of fig.4 is that $T_c$ decreases with Li doping for x=0.3 sample and absence of superconductivity for x=0.5 sample. Currently the BCS electron-phonon mechanism [3, 4] and the unconventional hole superconductivity mechanism [5, 6] have been proposed to explain the high temperature superconductivity observed in $MgB_2$. The two theories predict different behaviors of doping dependence of $T_c$ for $Mg_{1-x}Li_xB_2$. It has been estimated that the total number of holes per B atom in $MgB_2$ is approximately 0.067 [6]. With Li doping, the total number of holes per B atom increases due to hole doping. The BCS electron-phonon mechanism expects $T_c$ to increase upon Li doping, while the unconventional hole superconductivity mechanism predicts $T_c$ to decrease upon Li doping. Our results seem to support the hole superconductivity model.

Lorenz et al [7] has carried out the high pressure study on $MgB_2$ and $T_c$ of the samples were found to decrease linearly with the increase of the hydrostatic pressure. In this experiment, the hydrostatic pressure decreases both the in-plane and inter-plane B-B distance, leading to the decrease of $T_c$ for $MgB_2$. In our experiment, Li doping decreases the in-plane B-B distance and the inter-plane B-B distance remains unchanged, and $T_c$ also decreases with Li doping. This comparison suggests that the in-plane B-B distance is important for high $T_c$ of $MgB_2$ and the decrease of this distance leads to the decrease of $T_c$. It is not clear whether $T_c$ can be increased if the in-plane B-B distance is increased to some extent.

There are some other reports about the doping effect of $Mg_{1-x}M_xB_2$ (M=Al, Be) [8, 9]. For Al doping [8], $T_c$ decreases with the increase of the dopant concentration. Both lattice parameters *a* and *c* decrease with doping, in contrast to the behavior we observed in $Mg_{1-x}Li_xB_2$. For Be doping [9], it was found that Be can' t dope into Mg site and $T_c$ of $Mg_{1-x}Be_xB_2$ has the same $T_c$ as pure $MgB_2$.

In summary, nearly single phased $Mg_{1-x}Li_xB_2$ samples were prepared by solid state reaction method for $x\leq 0.3$. Li doping decreases the in plane lattice parameter *a*, which is proportional to the B-B distance in the B layer and does not have obvious effect on the lattice parameter *c*, which is proportional to the distance between the B layers. Li doping also decreases $T_c$ of $Mg_{1-x}Li_xB_2$ and superconductivity loss occurs for x=0.5 sample. The results of our work favor the hole superconductivity mechanism.

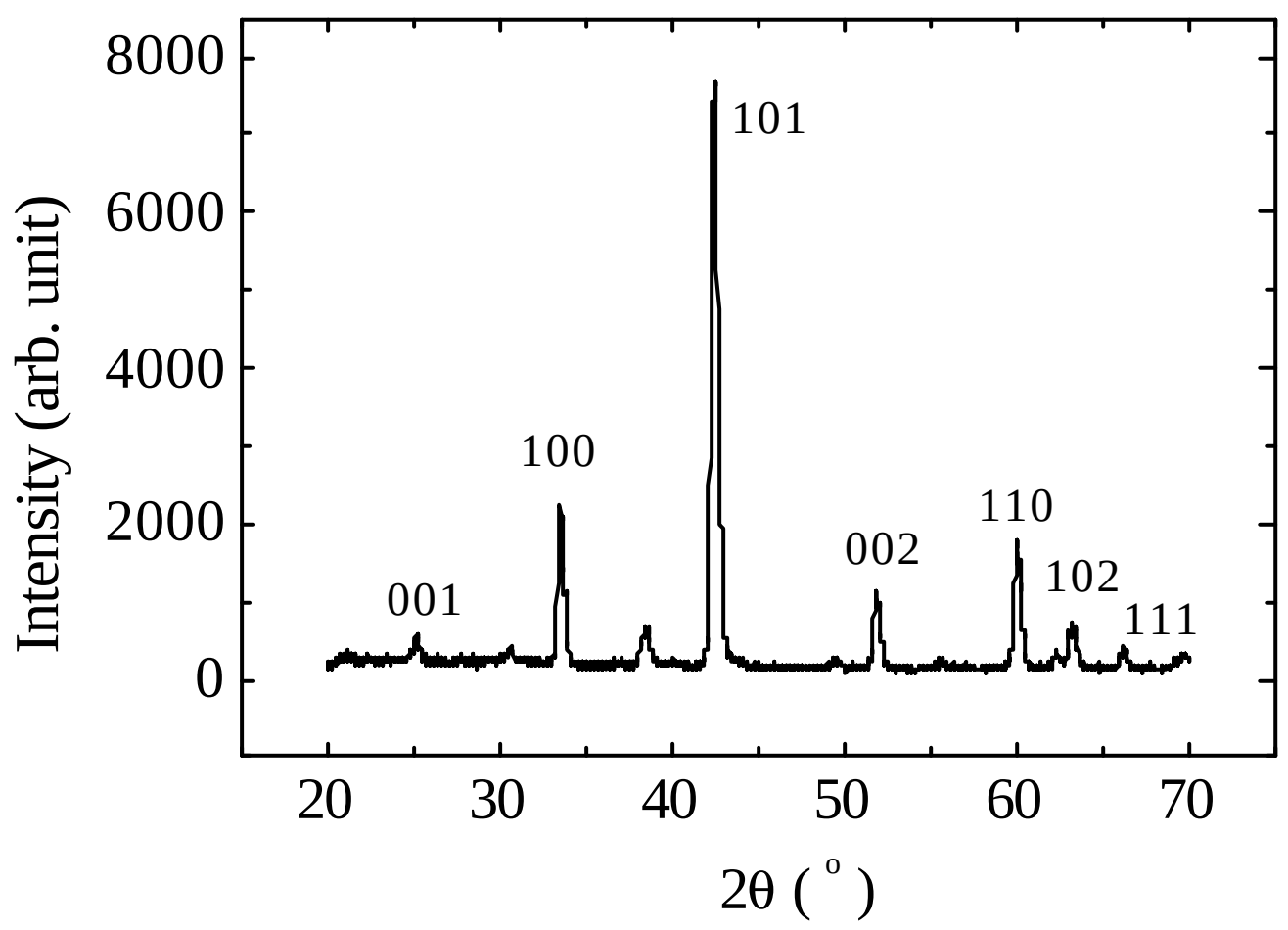

Fig. 1 x-ray diffraction pattern for $Mg_{0.9}Li_{0.1}B_2$.

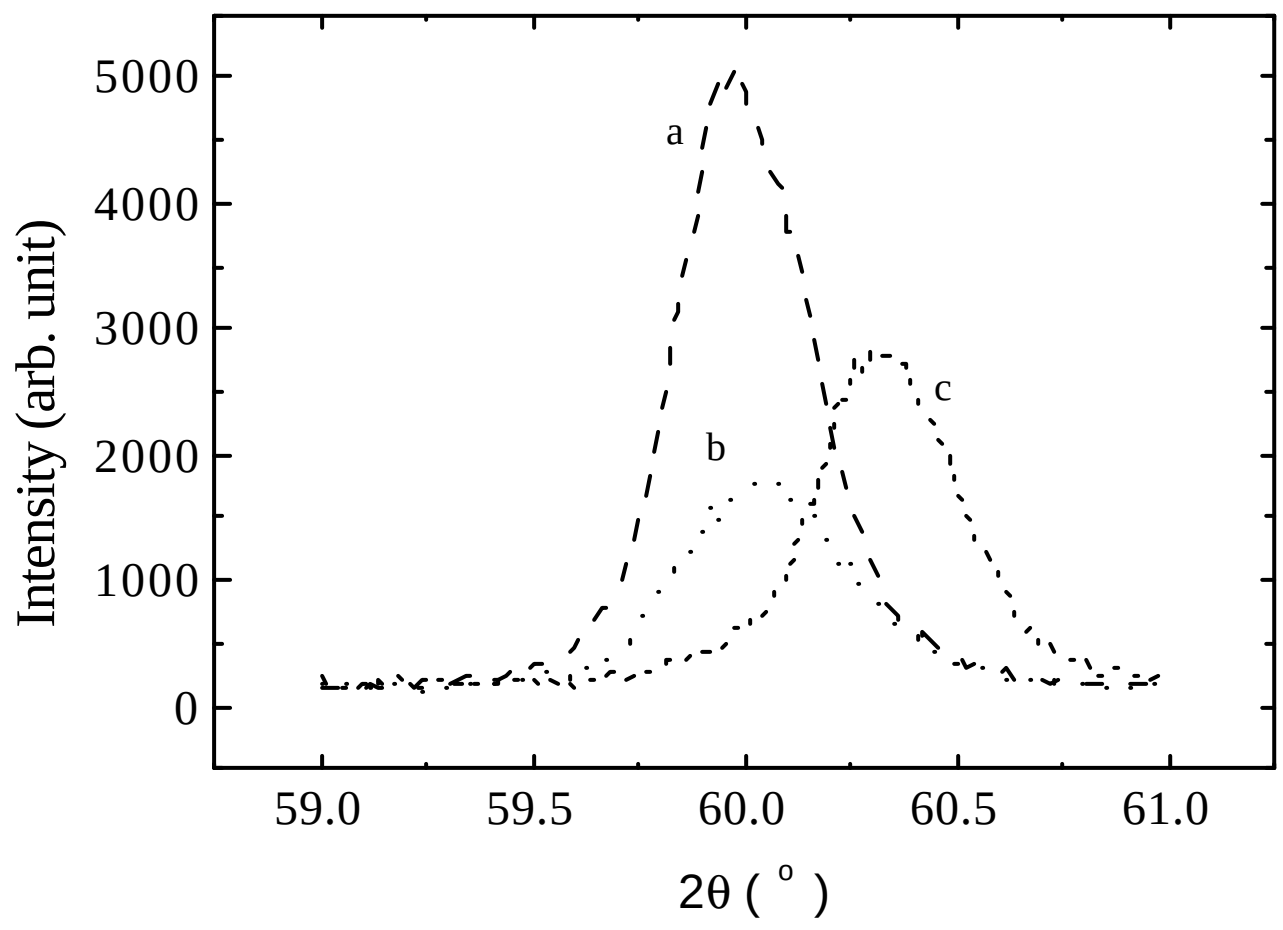

Fig. 2 Variation of the peak position for (110) peak of $Mg_{1-x}Li_xB_2$ with (a) x=0 (b) x=0.1 (c) x=0.3.

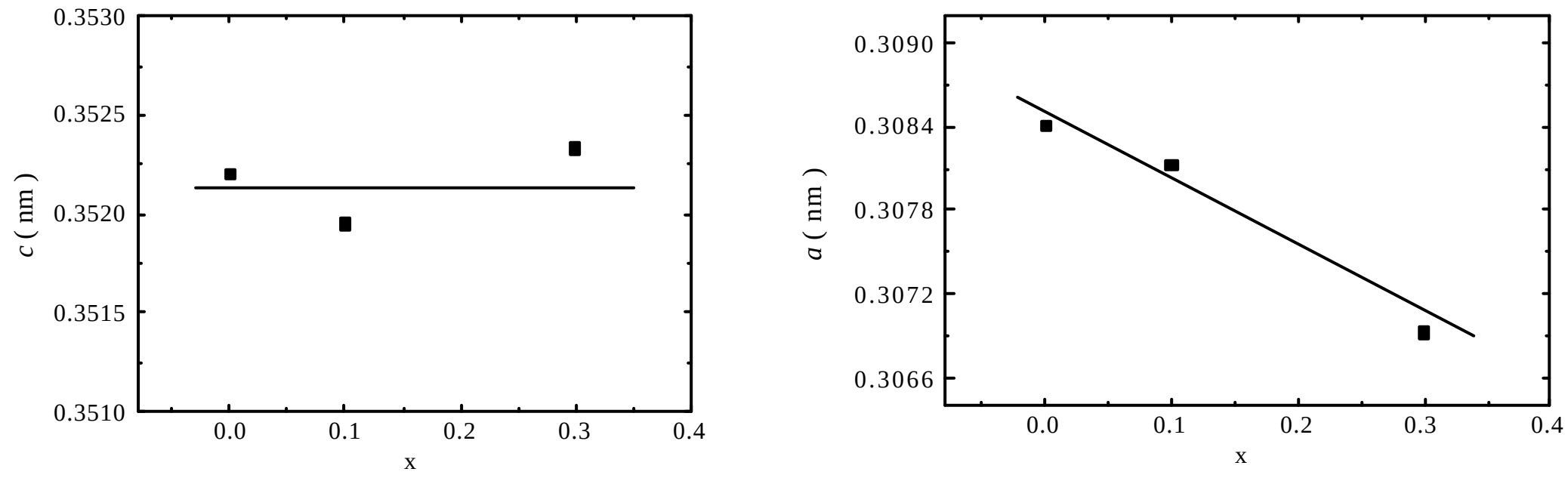

Fig. 3 Variation of the lattice parameters *a* and *c* with Li doping.

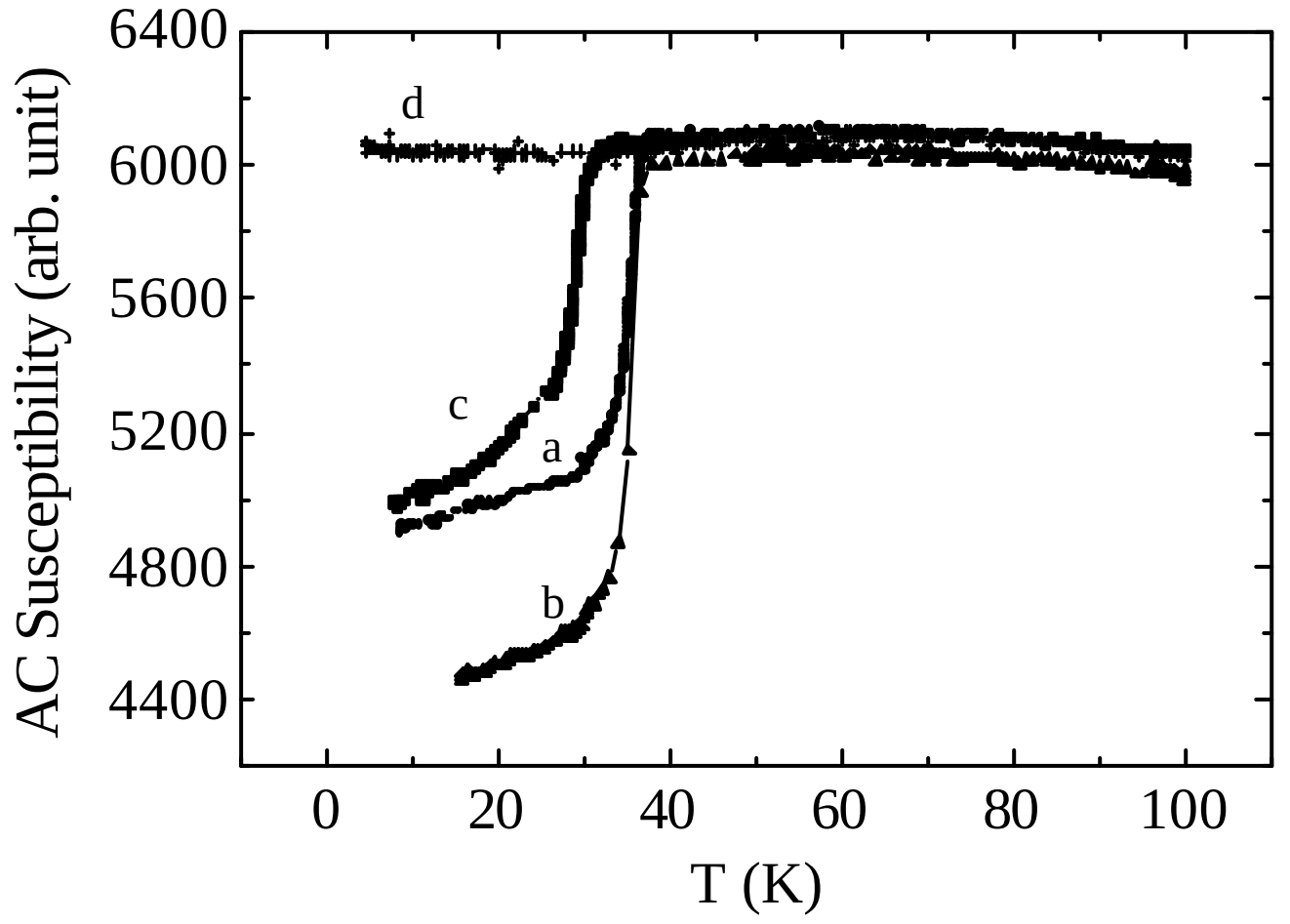

Fig. 4 Temperature dependence of AC susceptibility for $Mg_{1-x}Li_xB_2$ with different Li doping, (a) x=0 (b) x=0.1 (c) x=0.3 (d) x=0.5